\documentclass[12pt]{article}
\usepackage{epsf}
\title{{\bf Calculation of the pentaquark width by  QCD sum rule}}

\author{A.G.Oganesian\\
Institute of Theoretical and Experimental Physics,\\
B.Cheremushkinskaya 25, 117218 Moscow,Russia}
\date{}

\begin{document}

\maketitle

\newcommand{\be}{\begin{equation}}
\newcommand{\ee}{\end{equation}}

\def\la{\mathrel{\mathpalette\fun <}}
\def\ga{\mathrel{\mathpalette\fun >}}
\def\fun#1#2{\lower3.6pt\vbox{\baselineskip0pt\lineskip.9pt
\ialign{$\mathsurround=0pt#1\hfil##\hfil$\crcr#2\crcr\sim\crcr}}}

\vspace{1cm}
\begin{abstract}

The pentaquark width is calculated in QCD sum rules. Result for
$\Gamma_{\Theta}$ show, that $\Gamma_{\Theta}$ can vary in the
region less than 1$MeV$. The main conclusion is, that if
pentaquark is genuine states then sum rules really predict the
narrow width of pentaquark $\theta^+$, and the suppression of the
width is both parametrical and numerical.

\end{abstract}


\vspace{1cm}

\normalsize

In this paper we will discuss the the narrow exotic baryon
resonance $\Theta^+$ with quark content $\Theta^+ = uudd\bar{s}$
and mass 1.54 GeV. This resonance had been discovered  by two
groups \cite{n1,n2}. Later, the existence of this resonance was
confirmed by many other groups, although some searches for it were
unsuccessful. Moreover, last year some groups, which have seen
pentaquark, in the new experiments with higher statistics.
reported null result for pentaquark signal (CLAS experiments on
hydrogen and deuterium $\cite{ex1}$, BELLE $\cite{ex2}$) but at
the same time DIANA $\cite{dolg}$, and also LEPS,  SVD-2 confirm
their results with higher statistic(see $\cite{buk}$ for the
review).

So the status  of $\Theta^+$, predicted in 1997 by D.Diakonov,
V.Petrov and M.Polyakov \cite{n4} in the Chiral Soliton Model,
till now is doubtful. (Some theoretical reviews are given in
\cite{n5,n6}).

One of the most interesting features of $\Theta^+$ is its very
narrow width. Experimentally, only an upper limit was found, the
stringer bound was presented in \cite{n2}: $\Gamma < 9 MeV$. The
phase analysis of $KN$ scattering results in the even stronger
limit on $\Gamma$ \cite{n7}, $\Gamma < 1 MeV$. A close to the
latter limitation was found in \cite{n8} from the analysis of $Kd
\to ppK$ reaction and in \cite{n9} from $K+Xe$ collisions data
\cite{n2}. Also $\cite{ex2}$ from the negative result of the
experiment give the upper limit for pentaquark width less than
640$KeV$

Such extremely narrow width of $\Theta^+$ (less than $1 MeV$)
seems to be very interesting theoretical problem. In the paper
\cite{ja1}, \cite{ja2} it was shown, that if pentaquark is genuine
state it width should be strongly parametrically suppressed. It
was shown, that the conclusion does not depend of the choice of
the pentaquark current (without derivatives). Recently in
\cite{niels} the numerical estimation of pentaquark width  (about
0.75 $Mev$ for pentaquark state with positive parity) was obtained
also by use QCD sum rules. The main goal of this paper is to find
numerical estimation of pentaquark width by use the method offered
in \cite{ja1},\cite{ja2}.

 \bigskip

{\bf \large Part 1. } In the papers \cite{ja1}, \cite{ja2} it was
shown, that pentaquark width should be suppressed as
$\Gamma_{\Theta} \sim \alpha^2_s \langle 0 \vert \bar{q} q \vert 0
\rangle^2$,  (for any current without derivatives). Let us shortly
remind the main points of the method. We start from 3-point
correlator

\be
 \Pi_{\mu}=\int e^{i(p_1x-qy)} \langle 0\mid
\eta_{\theta}(x)j^5_{\mu}(y) \eta_n(0)\mid 0 \rangle \ee

where $\eta_n(x)$ is the neutron quark current \cite{nbl1},
($\eta_n =\varepsilon^{abc} (d^a C \gamma_{\mu} d^b) \gamma_5
\gamma_{\mu} u^c $),

$ \langle 0\mid \eta_n\mid n\rangle =\lambda_n v_n$,
($v_n$ is the  nucleon spinor), $\eta_{\theta}$ is an arbitrary
pentaquark current
 $ \langle 0\mid \eta_{\theta}\mid \theta^+
\rangle=\lambda_{\theta} v_{\theta}$ and $j_{\mu 5} = \bar{s}
\gamma_{\mu}\gamma_5 u$ is the strange axial current.

As an example of $\eta_{\theta}$ one can use the following one
(see \cite{jap}, where it was first offered, and also \cite {ja3},
where the sum 2-point rule analysis for this current was
discussed):

\be
J_A =\varepsilon^{abc} \varepsilon^{def} \varepsilon^{gcf}
(u^{a^T} Cd^b ) (u^{d^T} C\gamma^{\mu} \gamma_5 d^e)\gamma^{\mu}
c\bar{s}_g\ee

and we will use it farther to obtain numerical results.

As usual in QCD sum rule the physical representation of correlator
(1) can be saturated by lower resonance states plus continuum
(both in $\eta_{\Theta}$ and nucleon channel)
\be
 \Pi^{Phys}_{\mu}=\langle 0\mid \eta_{\theta}\mid \theta^+ \rangle
\langle \theta^+ \mid j_{\mu}\mid n \rangle \langle n\mid \eta_n
\mid 0 \rangle \frac{1}{p^2_1 -m^2_{\theta}}\frac{1}{p^2_2-m^2}+
cont. \ee

where $p_2=p_1-q$ is nucleon momentum, $m$ and
$m_{\theta}$ are nucleon and pentaquark masses.

Obviously, in the limit of massless kaon \be \langle \theta^+ \mid
j_{\mu}\mid n \rangle =g^A_{\theta n} \bar{v}_n \Biggl (g^{\mu\nu}
-\frac{q^{\mu}q^{\nu}}{q^2}\Biggr ) \gamma^v\gamma_5 v_{\theta}
\ee

where axial transition constant $g^A_{\theta n}$ is just we are
interesting in (the width is proportional to the square of this
value). Such a method for calculation the width in QCD sum rules
is not new, see, e.g. \cite{nblk0}. In the case of massive kaon
the only change is in denominator of second term in r.h.s of the
eq. (4), i.e. \be \langle \theta^+ \mid j_{\mu}\mid n \rangle
=g^A_{\theta n} \bar{v}_n \Biggl (g^{\mu\nu}
-\frac{q^{\mu}q^{\nu}}{q^2-m_k^2}\Biggr ) \gamma^{\nu}\gamma_5
v_{\theta} \ee It is clear that the second term vanishes at small
$q^2$.

Substituting  $ \langle 0\mid \eta_n\mid n\rangle =\lambda_n v_n$,
 and  $ \langle 0\mid \eta_{\theta}\mid \theta^+
\rangle=\lambda_{\theta} v_{\theta}$  in eq (3) and take the sum
on polarization one can easily see, that (in the limit of small
$q^2$) correlator (1) is proportional to $g^A_{\theta n}$.

\be
 \Pi^{Phys}_{\mu}=\lambda_n\lambda_{\theta}g^A_{\theta n}
\frac{1}{p^2_1
-m^2_{\theta}}\frac{1}{p^2_2-m^2}(-2\hat{p_1}\gamma_5 p_1^{\mu} +
 ....) \ee

where dots in r.h.s mean other kinematic structures (proportional
to $q$  e.t.c).
 For our sum rules we will use invariant amplitude just at the kinematical
 structure $\hat{p_1}p_1^{\mu}$,
because, as it was discussed in $\cite{nblk}$, $\cite{nbls}$,
$\cite{nblk1}$, $\cite{nblk2}$ the choice of the kinematic
structures with maximal number of momentum lead to better sum
rules.

By use of the equation of motions the eq.(4) close to the mass
shell can be rewritten \be \langle \theta^+ \mid j_{\mu}\mid n
\rangle =g^A_{\theta n} \bar{v}_n \Biggl (\gamma^{\mu}
+\frac{m_{\theta}+m_n}{q^2}q^{\mu}\Biggr )\gamma_5 v_{\theta} \ee

At the same time, the second term in  (4,7) correspond to the kaon
contribution to $\theta -n$ transition with lagrangian density
$L=ig_{\theta n k}v_{n}\gamma^5 v_{\theta} \phi_k$

so one can write

\be \langle \theta^+ \mid j_{\mu}^5\mid n \rangle =g_{\theta n k}
\frac{q^{\mu}f_k}{q^2-m_k^2}\bar{v_{n}}\gamma^5 v_{\theta}\ee

Comparing (7) and  (8) one can found  (if we for a moment neglect
the kaon mass)

\be g_{\theta n k}f_k=(m_n+m_{\theta})g_{\theta n}^A\ee

This is the analog of the Golderberger-Trieman relation. Of course
the accuracy of this relation is about the scale of SU(3)
violation but as estimation of the value of $g_{\theta n k}$ it is
enough good. Before we discuss numerical  sum rules for $g_{\theta
n}^a$, let us remind the common properties of correlator (1).   In
\cite{ja1}, \cite{ja2} it was shown, that correlator(1) vanishes
in the chiral limit for any pentaquark current without
derivatives, so axial constant  $g_{\theta n}^a$ should be
proportional to the quark condensate. In this papers also the
another reason of small value of $g_{\theta n}^a$ was discussed.
Let us again consider correlator (1). One can easily note, that
unit operator (as well as $d=3$ operator e.t.c.) contributions to
the correlator  are expressed in the terms of the following
integrals \be \int e^{i(p_1x-qy)} \frac{d^4 x d^4 y}{((x-y)^2)^n
(x^2)^m} \equiv \int \frac{e^{ip_1x}}{(x^2)^m} \frac{e^{-iq
t}}{(t^2)^{n}} d^4 xd^4t \ee

It is clear that such integrals have imaginary part on $p_2^2$ and
$q^2$ - the momentum of nucleon and axial current - but there is
no imaginary part on $p_1^2$ - the momentum of pentaquark. So we
come to the conclusion that such diagrams  correspond to the case,
when there is no $\Theta^+$ resonance in the pentaquark current
channel (this correspond to background of this decay). (Note, that
this conclusion don't depend on the fact that one of the quark
propagators should be replaced by condensate, as we discuss
before). The imaginary part on $p_1^2$ (i.e. $\Theta^+$
resonance)appears only if one take into account hard gluon
exchange. So we come to conclusion, that if $\Theta^+$ is a
genuine 5-quark state (not, say, the $NK$ bound state), then the
hard gluon exchange is necessary, what leads to additional factor
of $\alpha_s$. We see, that pentaquark width $\Gamma_{\Theta} \sim
\alpha^2_s \langle 0 \vert \bar{q} q \vert 0 \rangle^2$, i.e.,
$\Gamma_{\Theta}$ has strong parametrical suppression.

Now we want to check, does this parametrical suppression really
lead to numerical smallness of the pentaquark width. It is
necessary to note, that in the sense of discussion before (see
$\cite{ja2}$) it is quite necessary to keep only those part of the
correlator, which has really  imaginary part both on $p_1^2$ and
$p_2^2$ (i.e pentaquark and nucleon 4-momenta square). This mean,
at least, double Borel transformation (for $p_1^2$ and $p_2^2$
independently). From  (5) it is clear that for invariant amplitude
at the kinematic structure $\hat{p_1}p_{\mu}$ we have the
following sum rules

\be \lambda_n\lambda_{\theta}g^A_{\theta n}
e^{-(m_n^2/M_n^2+m_{\theta}^2/M_{\theta}^2)} = (-1/2)B_{\theta}B_n
\Pi^{QCD}\ee

where $B_{\theta}, B_n$ mean Borel transformation on pentaquark
and nucleon momenta correspondingly, and continuum extraction is
supposed.

For $\Pi^{QCD}$  (QCD representation of correlator (1))
corresponding diagrams are shown fig.1a,b. But as was discussed
before, diagrams at Fig.1a has no imaginary part on pentaquark
4-momentum. (They are proportional to terms as in eq. (10)). So
the only contribution come from diagrams on the Fig.1b The
calculation of this diagrams is technically enough complicated.
One should pay special attention to extract the terms, which have
no imaginary part on pentaquark 4-momenta correctly. We perform
calculation in the x-representation, using standard exponential
representation of propagators and the relation $B
e^{-bp^2}=\delta(b-1/M^2)$. In this short paper we do not stop on
the discussion of calculation features. Unfortunately we can not
write down pure analytical answer, because the final answer is
expressed in terms of a very large number of different double (and
ordinary) integrals and it total size is very large (about some
hundred terms). That's why we prefer at last stage make numerical
analysis of results. Of course we check, that all integrals
converge if $Q^2 (=-q^2)$ is not equal to zero. At the same time
we use in calculation the fact, that the ratio
$A1=M_n^2/M_{\theta}^2$, (where $M_n^2, M_{\theta}^2$ are nucleon
and pentaquark Borel masses) should be of order of ratio of the
corresponding mass square $m_n^2/m_{\theta}^2$, so we can threats
$A1$ as small parameter.

Really, from the sense of sum rule, it is clear that we can found
axial constant $g_{\theta n}^A$ only at $Q^2$ not close to zero
(about $1 Gev^2$ or higher). Of course, as was discussed before,
the result of QCD part (we will denote it as $R^{QCD}
=B_nB_{\theta}\Pi^{QCD}$) is proportional to quark condensate and
strong coupling constant $\alpha_s a$, where $ a = -(2 \pi)^2
\langle 0 \vert \bar{q} q \vert 0 \rangle$. As a characteristic
virtuality we chose the Borel mass for nucleon, but one should
note, that because $\alpha_s a^2$ do not depend on normalization
point, this choice is rather unessential. We use the value of
$\alpha_s a^2$=0.23 $Gev^6$, $\lambda^2_n*32{\pi}^4=3.2Gev^6$,
\cite{bl2} $\lambda^2_{\theta}*(4{\pi})^8=12Gev^{12}$, obtained
\cite{ja3}. The continuum dependence is weak, we use standard
value $s_0=1.5GeV^2$ for nucleon and $s_0=4.-4.5GeV^2$ for
pentaquark current \cite{ja3}.

On Fig.2 the value of axial constant $g_{\theta n}^A$, obtained
from sum rules, are shown at two values of $Q^2$: $Q^2=1Gev^2$,
$Q^2=2Gev^2$ as a function of Borel mass of nucleon (the Borel
mass of pentaquark is supposed to be $M_{\theta}^2=3M_n^2)$. One
can see that the Borel mass behavior of $g_{\theta n}^A$ is
reasonable. I should note, that this is only contribution of the
first non vanishing operator (dimension 3), and the calculation
the contribution of higher dimension operators are in progress,
but first estimations show, that they expected not to be extremely
large (we expect contribution no more than 40$\%$). On the Fig.3
the $Q^2$ dependence of $g_{\theta n}^A$ is shown (at
$M_n^2=1Gev^2$ and $M_{\theta}^2=3M_n^2$ -thin line, and the same
for$M_n^2=1.2Gev^2$ - thick line ).
 Really we are interest the value
$g_{\theta n}^A$ in the limit $Q^2 \to 0$, which can't be
calculated directly from S.R., obviously. But one can see from
Fig.3, that $Q^2$ behavior is found to be almost linear (which is
physically expected at small $Q^2$) so one can extrapolate it to
zero. Of course such an extrapolation is lead to large inaccuracy,
especially before we have no calculate the higher order
corrections and that's why don't know the region of $Q^2$ where
sum rules are reliable. We suppose them to be reliable starting
$Q^2=1Gev^2$, then we estimate averaged (on Borel mass) $g_{\theta
n}^A=0.02$ at $Q^2=0$ with inaccuracy about a factor two. The
other sources of inaccuracy are:

 a) the method itself (the accuracy is of the order
of the SU(3) violation)

b) possible contribution of higher dimension operators (according
the estimations, most likely can change the value about 30-50$\%$)

c) in the value of $\lambda_{\theta}$ (accuracy about 20-30$\%$)

d) inaccuracy  of sum rule approach, especially for pentaquark
case (see, for example, discussion in \cite{nar}), and also the
possible effects of the size of pentaquark.

For all this reason we clearly understand, that really we can
estimate only the order of magnitude of the value of axial
constant $g_{\theta n}^A(0)$ at $Q^2=0$, which can to be varied
from 0 to 0.06 with central point (formally) $g_{\theta
n}^A=0.02$. So we can more or less reliable estimate only upper
limit for $g_{\theta n}^A <0.06$. By use of eq. (9) one can easily
express the pentaquark width in terms of $(g_{\theta n}^A)^2$, and
we come to conclusion, that our result give upper limit for
pentaquark width $\Gamma_{\Theta} < 1Mev$. More precise prediction
can be done only after higher dimension operator corrections will
be estimated.  Note, that our result for width of the pentaquark
(with positive parity) does not contradict to those (0.75$MeV$),
obtained in \cite{niels} also in sum rules, but by quite different
method.

Our estimation of the pentaquark width  also doesn't contradict to
the result \cite {dolg} (0.36$MeV$ with accuracy about 30$\%$),
obtained from the ratio between numbers of resonant and
non-resonant charge exchange events.

The main conclusion is, that if the $\theta^+$ is genuine states
then  sum rules really predict the narrow width of pentaquark, and
the suppression of the width is both parametrical and numerical.

Author is thankful to B.L.Ioffe for useful discussions and
advises. This work was supported in part by CRDF grant
RUP2-2621-MO-04 and RFFI grant 06-02-16905.

\newpage

\begin{figure}
\epsfxsize=5cm \epsfbox{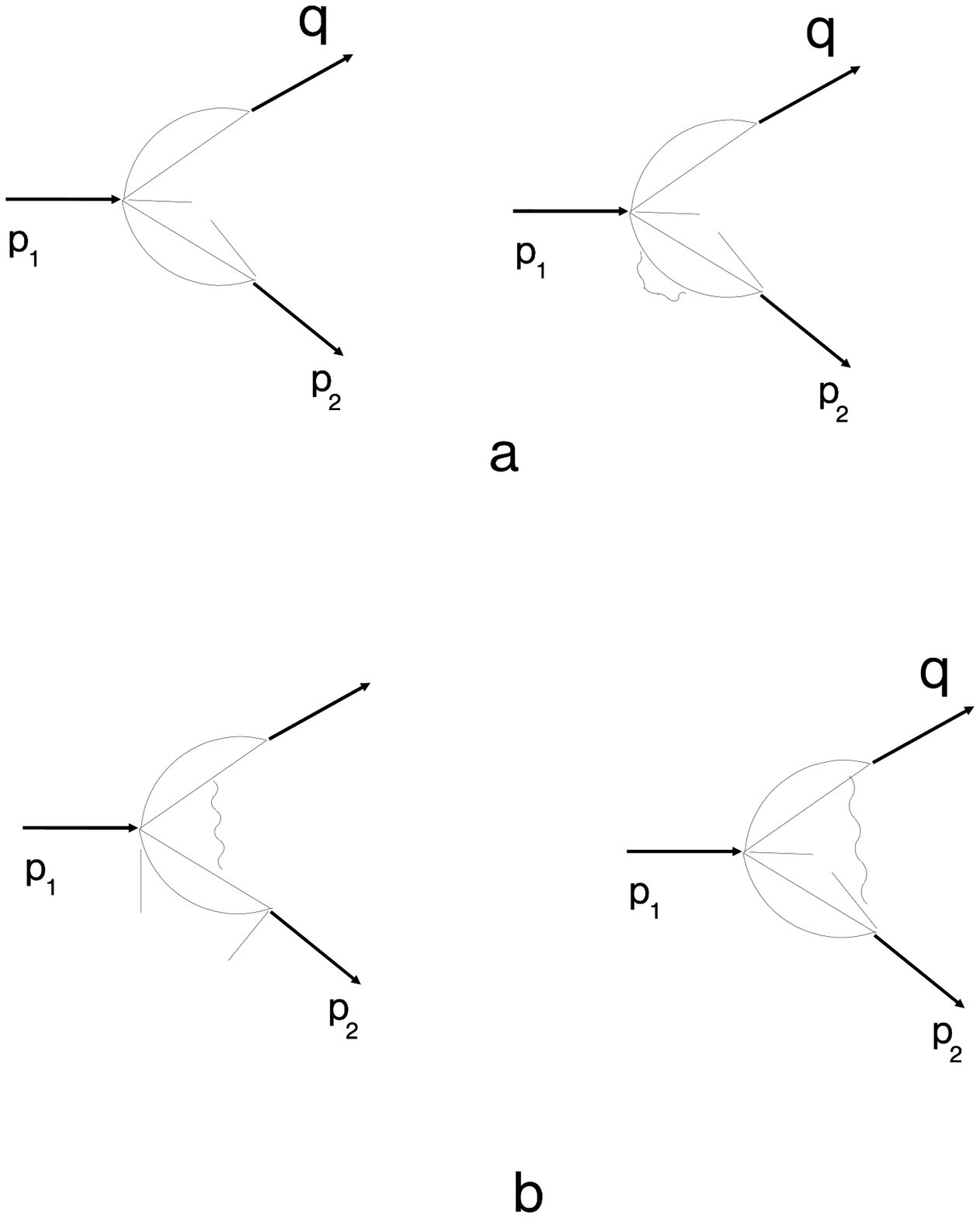} \caption{}
\end{figure}

\begin{figure} \epsfxsize=5cm \epsfbox{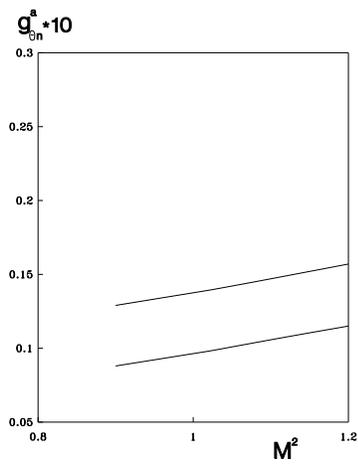}
\caption{$g^A_{\theta n}$ dependence on Borel mass for
$Q^2=1GeV^2$ - upper line and for $Q^2=2GeV^2$ - lower line}
\end{figure}

\begin{figure}
\epsfxsize=5cm \epsfbox{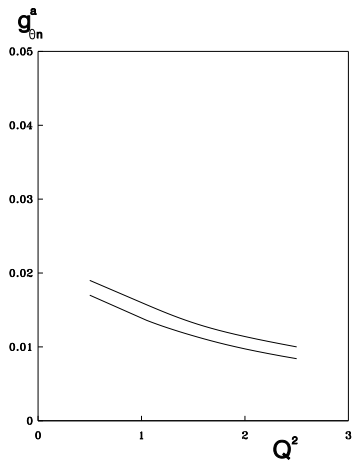} \caption{ $Q^2$ dependence of
$g^A_{\theta n}$ for$M_n^2=1.2GeV^2$ - upper
 line and for $M_n^2=1GeV^2$ - lower line}
\end{figure}

\end{document}